# A plasticity model with yield surface distortion for non proportional loading.

*M.L.M. FRANÇOIS*
*LMT Cachan (ENS de Cachan /CNRS URA 860/Université Paris 6)*

## ABSTRACT

In order to enhance the modeling of metallic materials behavior in non proportional loadings, a modification of the classical elastic-plastic models including distortion of the yield surface is proposed. The new yield criterion uses the same norm as in the classical von Mises based criteria, and a "distorted stress" $S_d$ replacing the usual stress deviator S. The obtained yield surface is then "egg-shaped" similar to those experimentally observed and depends on only one new material parameter. The theory is built in such a way as to recover the classical one for proportional loading. An identification procedure is proposed to obtain the material parameters. Simulations and experiments are compared for a 2024 T4 aluminum alloy for both proportional and nonproportional tension-torsion loading paths.

## ADDRESS

Marc François, LMT, 61 avenue du Président Wilson, 94235 Cachan CEDEX, France.

## E-MAIL

francois@lmt.ens-cachan.fr

## INTRODUCTION

Most of the phenomenological plasticity models for metals use von Mises based yield criteria. This implies that the yield surface is a hypersphere whose center is the back stress X in the 5-dimensional space of stress deviator tensors. For proportional loadings, the exact shape of the yield surface has no importance since all the deviators governing plasticity remain colinear tensors. In contrast, for non proportional loadings, the stress path may affect any point of the yield surface, and this shape becomes very important as its normal gives the direction of the plastic flow through normality (Hill [1950]).

Phillips and co-workers (Phillips & Gray [1961]; Phillips & Tang [1972]) carried out the first experiments on pure aluminum and obtained very complex shapes for yield surfaces: for each straight loading path, a corner forms in the front end and a flat zone forms in the rear. The model developed by Kurtyka & Zyczkowski [1985; 1996] is able to model such complex distortion with geometric considerations and material constants.

Many authors (Bui [1966]; Shiratori *et al.* [1976a, 1976b]; Winstone [1983]; Rousset [1985]; Rousset & Marquis [1985]; Cheng & Krempl [1991]; Wu & Yeh [1992]; Khan & Wang [1993]; Boucher *et al.* [1995]; Ishikawa [1997]) have also measured or modeled (Gupta & Meyers [1994]) distortion of yield surfaces. However, they have observed more "egg-shaped" surfaces (see Figs. 1 and 2). This simple shape will be our first assumption and will be more precisely defined later. The second hypothesis made here is that the egg-axis is the backstress X, in other words that the yield surface remains invariant respect to a rotation (in the deviatoric space) around X. Dahan *et al.* [1988] introduce a new internal variable to describe the distortion with (S-X) as the egg-axis (S being the stress deviator). The choice between these two directions cannot be made on the basis of available testing and the chosen direction seem to fit the experiments correctly. A third hypothesis is that the distortion from the sphere to the egg is proportional to the ratio of the norm of the backstress X to the new kinematic hardening limit $X_l$. This can be seen as a natural first step in modeling and leads to classical theory while $X_l$ is set at an infinite value.

The entire model is developed within the framework of irreversible thermodynamics (Haplhen & Nguyen [1975]; Lemaitre & Chaboche [1990]) involving the Helmotz free energy $\rho\Psi$, the yield function f and the dissipation pseudo-potential F. The formulation allows an easy implementation in computerized finite element codes for a low calculation cost. An implicit Newton scheme for stress driven transformations is described and has been programmed.

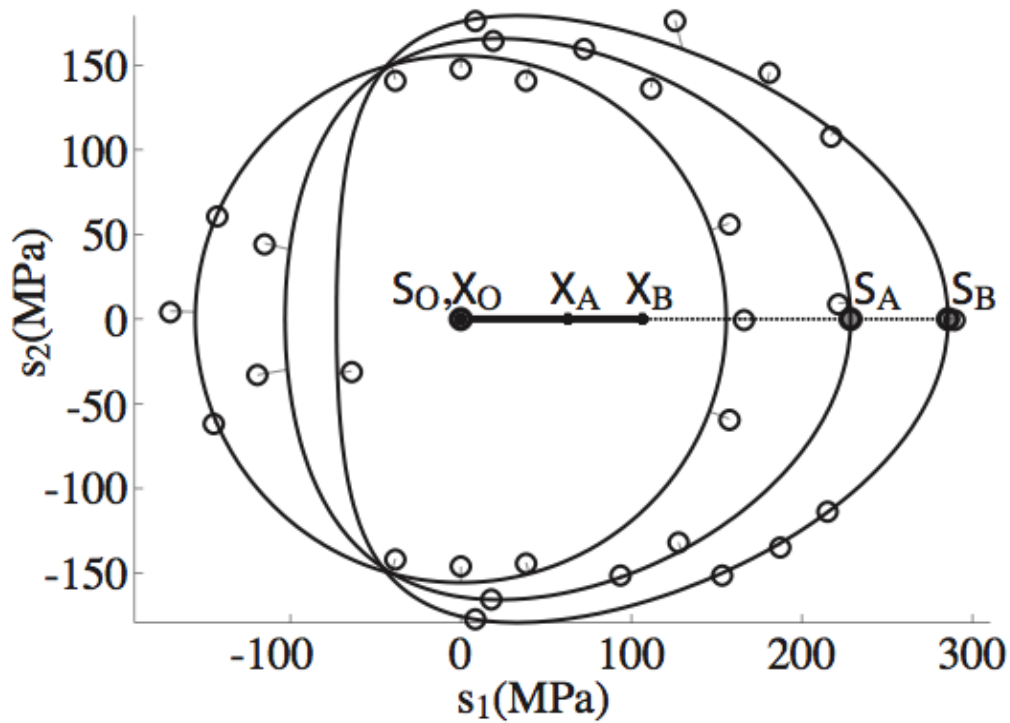

Fig. 1. Yield surfaces for monotonic loading.

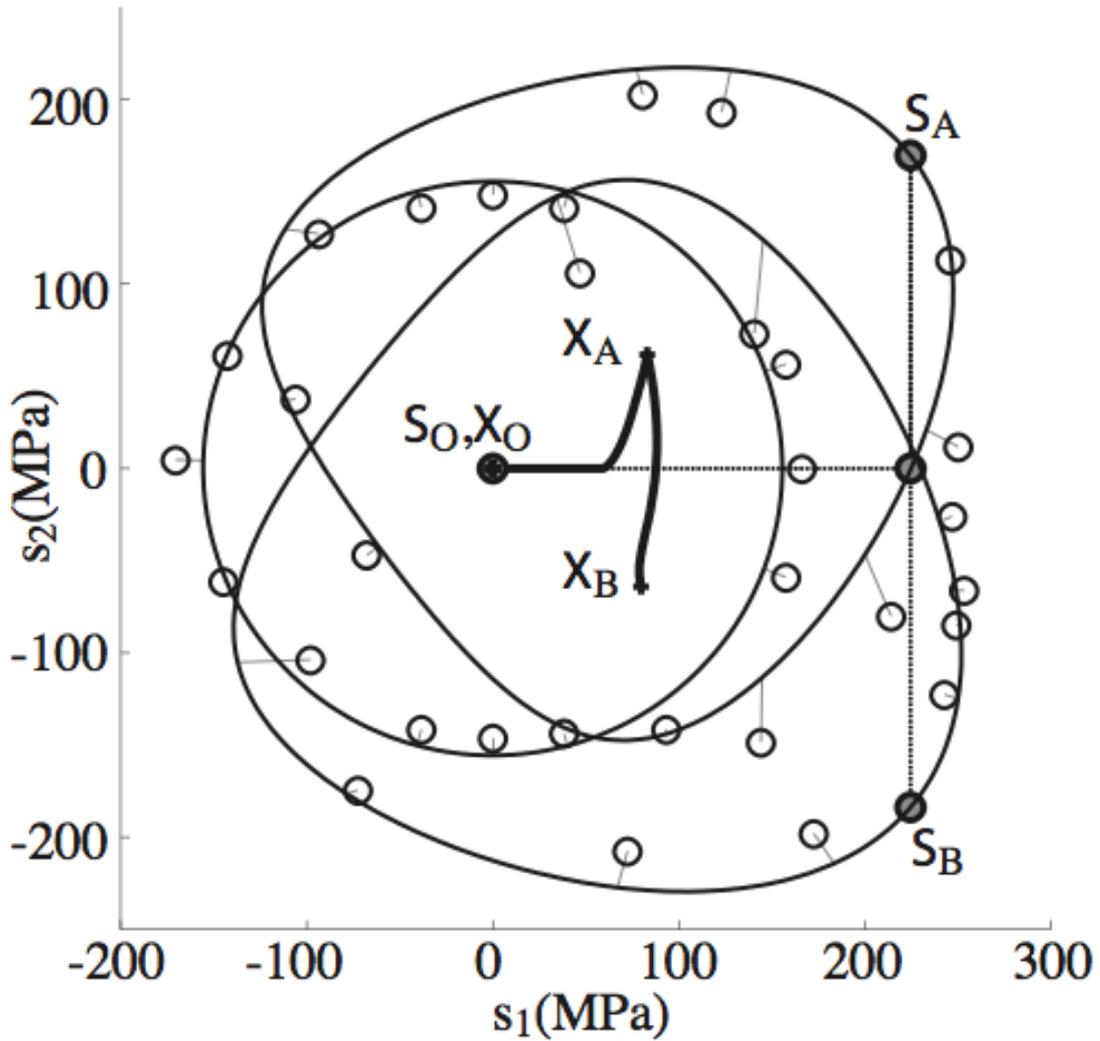

Fig. 2. Yield surfaces for non proportional tension-torsion testing.

# 1 CLASSICAL ELASTIC-PLASTIC MODEL

For details regarding the theory we refer to Lemaitre and Chaboche [1990]. Variables are the Cauchy's stress s associated with the elastic (reversible) strain $e_e$; the opposite of the stress deviator -S associated with the plastic (irreversible) strain $e_p$; the backstress X associated with the plastic strain a (kinematic hardening) and the isotropic hardening R associated with the plastic stain r (isotropic hardening). Constants are the stiffness tensor **K** and the plastic flow constant C. H(r) is the integral of the isotropic hardening function h(r). The thermodynamic forces $A_k$=(s, -S, X, R) are obtained by differentiation of the following Helmoltz free energy $\rho\Psi$ with respect to the state variables $V_k$=($e_e$, $e_p$, a, r).

$$\rho\Psi(e_e, a, r) = \frac{1}{2} e_e{:}\mathbf{K}{:}e_e + \frac{1}{2} C\, a{:}a + H(r) \qquad (1.1)$$

$$s = \mathbf{K}{:}e_e \qquad (1.2)$$

$$X = C\, a \qquad (1.3)$$

$$R = H'(r) = h(r) \quad (1.4)$$

The yield surface is described by the yield function f based on the von Mises normative expression. In this paper, all classical corrective terms in 3/2 or 2/3 (whose role is to simplify the equations obtained in tension-compression) have been removed as it has been preferred to simplify tensorial equations used in computerized calculation than some manual calculation used only for identifications. The yield stress $\sigma_y$ is defined then to be $\sqrt{2/3}$ of the classical value identified in tension-compression. This leads to description of the yield criterion as an hypersphere of radius $R+\sigma_y$.

$$f(S, X, R) = \| S - X \| - R - \sigma_y \quad (1.5)$$

If the plastic flow condition (François *et al.* [1991]) is satisfied, the following equations governs the plastic flow.

$$(f = 0) \ \& \ \left( \frac{\partial f}{\partial S} : dS > 0 \right) \Rightarrow (\text{plastic flow}) \quad (1.6)$$

The dissipation pseudo-potential F differs from the yield function f (Eqn 1.5) by a quadratic term in X which provides the non-linear kinematic hardening (Armstrong & Frederick [1966]) weighted by the corresponding material constant γ.

$$F(S, X, R) = f(S, X, R) + \frac{\gamma}{2C} X:X \quad (1.7)$$

The state variable evolution $dV_k$ is given by derivation of the dissipation pseudo-potential with respect to the associated thermodynamic forces $A_k$ and the plastic multiplier dλ.

$$dV_k = - \frac{\partial F}{\partial A_k} \ d\lambda \quad (1.8)$$

We deduce from Eqn (1.8) Hill's normality rule [1950] when considering the thermodynamic force -S (Eqn 1.9). The following flow rules take into account Eqn (1.7) in order to refer only to the yield function f instead of F.

$$de_p = \frac{\partial f}{\partial S} \ d\lambda \quad (1.9)$$

$$da = - \frac{\partial f}{\partial X} \ d\lambda - \frac{\gamma}{C} X \ d\lambda \quad (1.10)$$

$$dr = - \frac{\partial f}{\partial R} \ d\lambda \quad (1.11)$$

The value of the plastic multiplier dλ is given by the consistency equation df=0.

$$d\lambda = \left( C \frac{\partial f}{\partial X} : \frac{\partial f}{\partial X} + \gamma X : \frac{\partial f}{\partial X} + h'(h^{-1}(R)) \left( \frac{\partial f}{\partial R} \right)^2 \right)^{-1} \frac{\partial f}{\partial S} : dS \quad (1.12)$$

The following paper presents a modification of this theory involving only the yield function (Eqn 1.5). Others equations remain unaltered.

# 2 EXPERIMENTS AND OBSERVATIONS ON THE YIELD SURFACES

The experimental results used in this paper have been reported by Rousset [1985]. These involved 2024-T4 aluminum alloy using tubular thin-walled specimens allowing two dimensional loading in tension and torsion. The tensile stress deviator is either classically written in the canonic base as $\sigma(\mathbf{e}_1 \otimes \mathbf{e}_1)$ or in the tensorial base of the deviators presented below as $s_1 e_1$. The shear deviatoric stress is either classically written in the canonic base as $\tau(\mathbf{e}_1 \otimes \mathbf{e}_2 + \mathbf{e}_2 \otimes \mathbf{e}_1)$ or in the tensorial base of the deviators presented below as $s_2 e_2$. Correspondence between $\sigma$ and $s_1$, $\tau$ and $s_2$ is given Eqn (2.3). We first sketch this tensorial base of deviators different from Ilyushin's one (Tanaka [1994]).

## 2.1 Tensorial base

Most of the measurements of a material testing are related to tensorial values, the stress S and the plastic strain $e^p$ in the case of plasticity. In our case of two dimensional loadings, one can plot (in 2 dimensions) two curves related to variables with different physical dimension (*i.e.* stress to strain) and two others related to variables of the same tensor (stress to stress; plastic strain to plastic strain).

The first way to plot such curves is the use of the components related to the canonic base of the tensors, in our example $\sigma_{11}$ to $\sigma_{12}$ for stresses, $\varepsilon^p_{11}$ to $\varepsilon^p_{12}$ for strains. This leads to an affinity in the representation of the yield surface: the spherical von Mises criterion appears as an ellipse and, in the general case, the affinity can be confused with the distortion.

The second way is to use classical correction factors: $(\sigma; \tau\sqrt{3})$ for stresses and $(\varepsilon^p; \gamma^p/\sqrt{3})$ for plastic strains (with $\sigma=\sigma_{11}$; $\tau=\sigma_{12}$; $\varepsilon^p=\varepsilon^p_{11}$; $\gamma^p=2\varepsilon^p_{12}$). These corrective terms lead to a circular representation of the von Mises criterion but are defined only for the tension or torsion loadings; for example, the superposition of a (non acting in plasticity) hydrostatic pressure leads to a modification of these corrective terms.

The third way recalled here is to use an orthonormal tensorial base (Mehrabadi & Cowin [1990]). The five tensors $e_1 \ldots e_5$ are deviatoric: $e_1$ is relative to a tension along the $\mathbf{e}_1$ axis, $e_2 \ldots e_5$ are pure shears, and $e_6$ is spherical. Their structure is given below with respect to the canonical base.

$$e_1 = \frac{1}{\sqrt{6}} \begin{bmatrix} 2 & 0 & 0 \\ 0 & -1 & 0 \\ 0 & 0 & -1 \end{bmatrix} \quad e_2 = \frac{1}{\sqrt{2}} \begin{bmatrix} 0 & 1 & 0 \\ 1 & 0 & 0 \\ 0 & 0 & 0 \end{bmatrix} \ldots$$

$$e_5 = \frac{1}{\sqrt{2}} \begin{bmatrix} 0 & 0 & 0 \\ 0 & 1 & 0 \\ 0 & 0 & -1 \end{bmatrix} \quad e_6 = \frac{1}{\sqrt{3}} \begin{bmatrix} 1 & 0 & 0 \\ 0 & 1 & 0 \\ 0 & 0 & 1 \end{bmatrix} \quad e_I : e_J = \delta_{IJ} \tag{2.1}$$

As $e_1 \ldots e_5$ is a base of the deviatoric subspace and $e_6$ is the hydrostatic base, we have the following properties (H means the hydrostatic part of the stress s):

$$s = S + H \quad ; \quad S = \sum_{i=1:5}(s:e_i)\, e_i \quad ; \quad H = (s:e_6)\, e_6 \tag{2.2}$$

The presented testings have been made in tension-torsion whose classical notations for stresses are $\sigma$ for $\sigma_{11}$ and $\tau$ for $\sigma_{12}$. The corresponding projections in the proposed base are noted $s_1 = s:e_1$ and $s_2 = s:e_2$ with the following correspondence:

$$s_1 = \sqrt{\frac{2}{3}}\ \sigma\ ;\ s_2 = \sqrt{2}\ \tau \qquad (2.3)$$

In the same way, classical notation for plastic strains are $\varepsilon^p$ for $\varepsilon^p_{11}$ and $\gamma^p$ for $2\varepsilon^p_{12}$. The corresponding projections in the proposed base are noted $e_1 = e^p:e_1$ and $e_2 = e^p:e_2$ with the following correspondence:

$$e_1 = \sqrt{\frac{3}{2}}\ \varepsilon^p\ ;\ e_2 = \frac{\gamma^p}{\sqrt{2}} \qquad (2.4)$$

Use of a tensorial base leads to consider the plot of two components of a unique tensor as the cross section of the tensorial space by a well chosen hyperplane (Rychlewski [1984]). All the properties valid in the space of deviators will remain in the 2-dimensional representation. For example, the von Mises based criterion used in the classical model will be represented as a circle whose radius is equal to $R+\sigma_y$, independent of the nature of the loading; an angle defined in the tensorial space (for exaple $\theta$ defined later in Eqn (3.4)) is measurable on a plot such as Figs. (1 and 2).

Combined with the simplified criteria as in Eqn (1.5) and later in Eqn (3.1) in which the corrective terms in $\sqrt{3/2}$ are suppressed these bases lead to many other simplifications. For example, the slope of the stress-strain curve (Fig. 4) at the beginning of plasticity is exactly C for any loading (not only the tensile one); the cumulative plastic strain p defined as $\int \sqrt{de^p:de^p}$, is exactly the length of the curve in the plastic strain to plastic strain graph.

## 2.2 Experimental results

The first experiment (Figs. 1 and 4) is a monotonic tensile one, stopped at four different steps (O, A, I, B) for the measurement of yield surfaces at the following stresses $\sigma$=(0, 280, 320, 350 MPa) or $s_1$=(0, 229, 261, 286 MPa) (Eqn 2.3). Only the steps (0, A, B) have been plotted for clarity.

The second experiment (Fig. 2) is a non proportional tension-torsion loading stopped at four steps (0, $0_1$, I, A, B) with the respective stresses $\sigma$=(0, 275, 275, 275, 275) MPa and $\tau$=(0, 0, 75, 120, -130) MPa or $s_1$=(0, 225, 225, 225, 225) MPa and $s_2$=(0, 0, 106, 170, -184) MPa (Eqn 2.3). Only the steps (0, A, B) have been plotted for clarity.

At each step, a "straight loading" (a straight line in the deviatoric stress space) was prescribed from the approximate center of the yield surface (approximating X) until the detection of a non linearity in the stress to strain curve; the corresponding offset (strain) was set at $5.10^{-5}$. For each state M, the loading stress $S_M$ is plotted by a dark circle; each $i^{th}$ experimental point of the yield surface $\underline{S}_{iM}$ for the state M is plotted as an open circle (Figs. 1 and 2). In order to minimize the effect of the accumulated plasticity involved in the detection of the yield surface, the loading path was designed such as two successive yield stresses be as far as possible in the stress space (*i.e.* describing a star for five measurements).

It can be easily seen that the classical models, in which the yield surface is described by a von Mises based criterion, leading to a circular representation in Figs. (1 and 2), deviate significantly from the shape determined by experiments.

## 3 NEW PLASTICITY MODEL FOR ELASTIC-PLASTIC METALS

In this section we propose a new yield criterion from which we deduce evolution laws. Its construction is illustrated in Fig. (3).

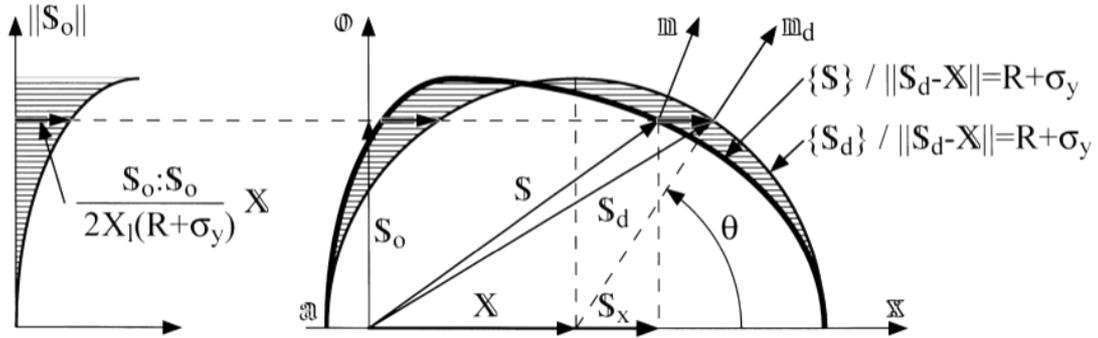

Fig. 3. Schematic of the yield surface.

In order to keep the very simple formulation given by the von Mises' norm, we introduce a new "distorted stress" $S_d$ which replaces the stress deviator S involved in the classical expression. One can remark that $S_d$ is not a new thermodynamic force, but only a function of existing variables. The other terms involved in the yield function f are the back stress X, the isotropic hardening function R and the yield stress $\sigma_y$ which is also the radius of the *initial* yield surface.

$$f(S, X, R) = \| S_d(S, X, R) - X \| - R - \sigma_y \quad (3.1)$$

The distorted stress $S_d$ is defined so that such as all the deviatoric stresses S for which the yield function f is null belong to a "hyper-egg" (first hypothesis presented in the introduction). We decompose the deviatoric stress S into its part $S_x$ collinear to X and it's orthogonal part $S_o$ according to:

$$S = S_x + S_o \quad ; \quad S_x = \frac{S:X}{\| X \|^2} X \quad (3.2)$$

We define the distortion to be parabolic (for simplicity) with respect to $\|S_o\|$ (can be seen in real size on Fig. 3).

$$S_d = S + \frac{S_o:S_o}{2 X_l (R + \sigma_y)} X \quad (3.3)$$

The loci of $S_d$ for f=0 correspond to a von Mises' hypersphere whereas the yield surface, *i.e.* the loci of S for f=0, correspond to an egg-shaped surface curve for any section by an hyperplane containing X (Fig. 3) and corresponds to a circle of radius $(R+\sigma_y)$ for any section by a hyperplane orthogonal to X. The egg's symmetry axis is the backstress X (second hypothesis). The distortion is linearly proportional to the ratio of the norm of X to the new constant introduced, the "kinematic hardening limit" $X_l$ (third hypothesis). This leads naturally to a virgin state (in which the back stress is null) satisfying the von Mises criterion and the existence of a limit state for which the distortion is complete. The distortion is maximum when $\|X\|=X_l$. At this step, the elastic domain has a locally flat zone (at stress a in Fig. 3), but remains convex as for all states. Note that the use of classical non linear kinematic hardening rules described in section (1) lead automatically to such limit value ($C/\gamma$) for $\|X\|$.

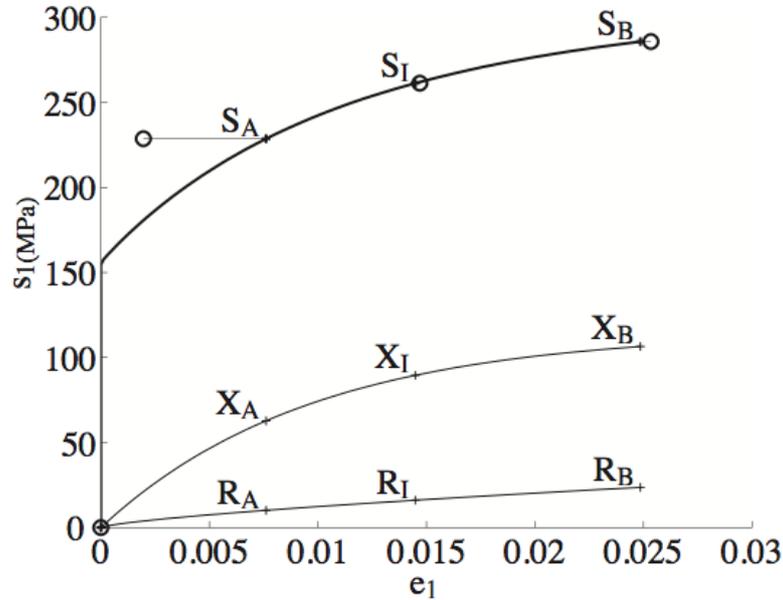

Fig. 4. Stress to plastic strain for tensile testing.

The backstress X represents, as for the classical reference model, the middle of the elastic domain in the case of proportional loading. Another choice would be to write the model in order to define X as the center of the maximum diameter of the egg but this leads to an influence of the distortion even on proportional loadings, making identification more difficult (François [1999]).

In proportional loadings, as tensorial variables (S, X, $e_p$) remain colinear, the othogonal part $S_o$ remains null, $S_d$ remains always equal to S and the classical model described in section (1) is recovered.

In the new model, the normal n of the yield surface is not collinear to S-X (as is the case for the classical model) when the loading is non proportional (when S does not remain collinear to X); this will affect behavior of the model during non proportional loadings involving plasticity, as in two-dimensional ratchetting for instance.

Another point of view can be given while using as parameter the following angle $\theta$ between ($S_d$ - X) and X represented in real size in Fig. (3) which allows us to describe every possible state for the material with a reduced set of variables:

$$\cos(\theta) = \frac{(S_d - X):X}{\|S_d - X\| \, \|X\|} \quad (3.4)$$

Let us suppose a state defined by (S, X, R). The tensors S and X define a hyperplane in the 5-dimensional space of deviators with an orthonormal basis x, o defined as:

$$x = \frac{X}{\|X\|} \quad ; \quad o = \frac{S_o}{\|S_o\|} \tag{3.5}$$

The yield surface in this plane can be described respect to the angle $\theta$ in the base $(x, o)$. The state is then fully described by X (the norm of X), R and $\theta$ (with $0 \leq \theta \leq \pi$). Let us consider now a case involving plasticity, *i.e.* f=0 (Eqn 1.6). The following equations give a simple geometrical interpretation of the yield surface and show clearly the link to classical theory ($X_l \rightarrow \infty$ leads to circular yield surfaces).

$$S_d - X = (R+\sigma_y)(\cos(\theta) x + \sin(\theta) o) \tag{3.6}$$

$$S - X = (R+\sigma_y)\left(\left(\cos(\theta) - \sin^2(\theta)\frac{X}{2X_l}\right) x + \sin(\theta) o\right) \tag{3.7}$$

Introducing the normal of the yield surface relative to the distorted stress $n_d$ (Fig. 3), the following gradients can be obtained with respect to the thermodynamic forces (-S, X, R):

$$n_d = \frac{\partial f}{\partial S_d} = \frac{S_d - X}{\|S_d - X\|} \tag{3.8}$$

$$\frac{\partial f}{\partial S} = n_d + \frac{n_d:X}{X_l (R+\sigma_y)} S_o$$

$$\frac{\partial f}{\partial X} = -\left(1 - \frac{S_o:S_o}{2X_l (R+\sigma_y)}\right) n_d - \left(\frac{n_d:S_x}{X_l (R+\sigma_y)}\right) S_o$$

$$\frac{\partial f}{\partial R} = -\left(\frac{S_o:S_o}{2X_l (R+\sigma_y)^2}\right) n_d:X - 1 \tag{3.9}$$

It can be again easily verified that the form of the classical model can be found either when $S_o$ is null (uniaxial proportional case) or when $X_l$ tends to infinity. All these equations can be rewritten with the angle $\theta$ (Eqn 3.4) in the $(x, o)$ base.

$$n_d = \cos(\theta) x + \sin(\theta) o \tag{3.10}$$

$$\frac{\partial f}{\partial S} = \cos(\theta) x + \sin(\theta)\left(1+\cos(\theta)\frac{X}{X_l}\right) o$$

$$\frac{\partial f}{\partial X} = -\left(1-\frac{(R+\sigma_y)\sin^2(\theta)}{2X_l}\right) \cos(\theta) x$$

$$-\left(1 + \frac{R+\sigma_y}{2X_l}(3\cos^2(\theta)-1) + \frac{X \cos(\theta)}{2X_l}\left(2-\sin^2(\theta)\frac{R+\sigma_y}{X_l}\right)\right) \sin(\theta) o$$

$$\frac{\partial f}{\partial R} = -\frac{\sin^2(\theta) \cos(\theta) X}{2X_l} - 1 \tag{3.11}$$

In the classical model, dr is always equal to dp, the increment of accumulated plastic strain p defined as $dp=\|de_p\|$. Fig. (5) shows the ratio dp/dr with respect to the angle $\theta$ in the case $\|X\|=X_l$ for which it's minima/maxima are the farthest from 1 but dp/dr always remains between 0.811 and 1.190 and is equal to 1 for proportional loadings ($\theta=0$ or $\theta=\pi$) and for $\theta=\pi/2$. This indicates that we can keep the constants of the isotropic hardening function h(r) unchanged from the ones identified for the classical model and one can say that the isotropic hardening is still driven by the accumulated plastic strain.

We have not yet proven analytically the positiveness of the dissipation dD:

$$dD = S:de_p - X:da - R\, dr$$

$$dD = \left( S:\frac{\partial f}{\partial S} + X:\frac{\partial f}{\partial X} + \frac{\gamma}{C} X:X + R\, \frac{\partial f}{\partial R} \right) d\lambda \qquad (3.12)$$

We use here a numerical study. The evolution equation (Eqns 1.9, 1.10 and 1.11) and the expression of gradients (Eqn 3.11) show that all the involved tensors are in the (x, o) plane even if dS is out of plane since only its projection is used in the expression of d$\lambda$ (Eqn. 1.12). The plastic multiplier d$\lambda$ is calculated while setting $\partial f/\partial S:dS$ to an arbitrary positive value (Eqn.1.6). The minimum of dD in the (x, o) plane is numerically determined with respect to a reduced set of variables [$\theta$, X, R] and material constants [$\sigma_y$, $X_l$, C, $\gamma$] (and the 2 constants [k, m] involved in the isotropic hardening power law h(r)=k $r^{1/m}$ used here). By normalizing by $\sigma_y$ all the parameters proportional to a stress (*i.e.* X, R, $\sigma_y$, $X_l$, C and k), we consider 8 parameters. For each of them, two values are considered, namely the minimum and maximum values: ($0 \leq \theta \leq \pi$), ($0 \leq X < C/\gamma$), ($0 \leq R < L\sigma_y$), ($C/\gamma \leq X_l < L\sigma_y$), ($0 < C < L\sigma_y$), ($0 \leq \gamma < L$), ($0 \leq k < L\sigma_y$), ($0 < m < L$), where L is a "large" value ($10^5$). To avoid reaching local minima, we carried out the minimization with $2^8$ initial values corresponding to the combinations for the 8 parameters with respect to their minimum / maximum values. Within these bounds, the minimum of the dissipation has been found positive. Therefore no new restriction on the choice of material constants is found to fulfill the second principle of thermodynamics.

An implicit scheme of Newton-Raphson type has been developed in order to model a stress path on a representative volume element. Calculation is then simple as the dissipation equations are stress driven and as elastic strain and Young's modulus are not involved. The Figs. (1 and 2) show in plain lines the simulated yield surfaces for each state, obtained with the material constants identified in section 4.

## 4    IDENTIFICATION

The material constants to be identified are the yield stress $\sigma_y$, the non linear kinematic hardening parameters (C, $\gamma$), the constant $X_l$ associated to the distortional effect and the isotropic hardening function h(r). The optimization procedure is first made using yield surface measurements (in the stress space) and then using the stress-strain curve. The experiment used for identification is the monotonic tensile test (Figs. 1 and 4).

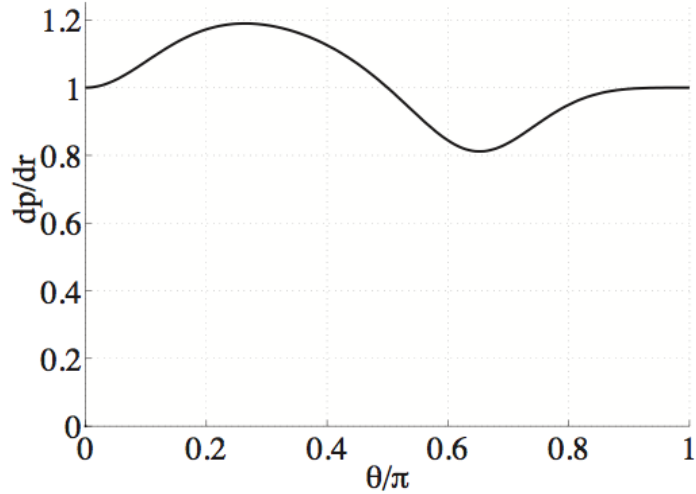

Fig. 5. Ratio dp/dr as a function of $\theta$.

We denote $\underline{S}_{iM}$ the $i^{th}$ experimentally measured stress belonging to the yield surface for the state M. The yield stress $\sigma_y$ has been identified as the average norm of measured yield surface stresses $\underline{S}_{iO}$ on virgin material (O denotes the virgin state, and the brackets denotes the average on the term i):

$$\sigma_y = <\|\underline{S}_{iO}\|> \tag{4.1}$$

Let us introduce the following distance $\underline{d}_M$ from the measured stresses $\underline{S}_{iM}$ to any theoretical yield surface defined by X and R at state M as follows:

$$\underline{d}_M(X, R; X_l, \sigma_y) = <|f(\underline{S}_{iM}, X, R; X_l, \sigma_y)|> \tag{4.2}$$

This definition is close (equal if there is no distortion) to the average of real distances (in the stress space) from $\underline{S}_{iM}$ to the yield surface defined by $(X, R; X_l, \sigma_y)$. As the considered tensile testing (used for identification) imposes that X is collinear to the stress deviator $e_1$ (Eqn 2.1), only the norm $X=\|X\|$ remains unknown. For each step M, the measured yield stresses $\underline{S}_{iM}$ allow us to determine numerically the three values $(X_M, R_M, X_{lM})$ which define the closest theoretical yield surface to experimental data (the closest egg) as follows:

$$(X_M, R_M, X_{lM}) = \arg\left(\min_{(X, R, X_l)}\left(\underline{d}_M(Xe_1, R; X_l, \sigma_y)\right)\right) \tag{4.3}$$

The kinematic hardening limit $X_l$ is chosen as the average of the $X_{lM}$ values obtained for each step. As the length of the egg-shaped yield surface along the $e_1$ axis is equal to $2(\sigma_y+R_M)$ in tension, the $R_M$ can be plotted versus the experimentally obtained accumulated plastic strain $p_M=\|\underline{e}_{pM}\|$ in our monotonic tension. This allows us to fit the isotropic hardening rule $R=h(r)$ (as $r=p$ in the considered monotonic case, see section 3). In this paper we have used a classical power law, $R=kr^{(1/m)}$ which needs at least three steps to be identified. The remaining C and $\gamma$ constants are identified with the stress versus plastic strain curve (Fig. 4); C is the slope of the curve at the beginning of plasticity and $\gamma$ is defined by the saturation value for X, *i.e.* $C/\gamma$. One can remark that the "natural value" for the kinematic hardening limit introduced, $X_l$ is $X_l=C/\gamma$. In the example, $X_l=130$ MPa and $C/\gamma=114.3$ MPa, this means that the complete distortion (locally flat back) cannot be reached. The values obtained for 2024-T4 aluminum alloy are given for a power isotropic hardening law $R=kr^{(1/m)}$ in Table (1).

## 5 COMPARISON OF THE SIMULATIONS WITH THE EXPERIMENTS

Let us compare theoretical (solid lines) to experimental (circles) results for the monotonic tensile test described in section 3. The constants used (Table 1) have been obtained from the identification process described above. Concerning the stress versus plastic strain curve (Fig. 4), the first step A is not accurately fit, due to the simple hardening rules used (the distortion modification is not involved in this curve for monotonic loading). Fig. (1) allows the comparison of yield surfaces in the tension-torsion [ $e_1$ $e_2$ ] hyperplane of deviatoric stresses (according to tensorial base, Eqn 2.1). The average distance (Eqn 4.2) between simulation and measurements is 5.99 MPa for the four steps. The same distance, computed with the same material constants but with $X_l$ set to an infinity (classical model) is 11.18 MPa. In this case, the yield surface could be represented (on Fig. 1) by a circle coinciding to the egg-shape of the presented model at the stresses collinear to X (see Fig. 3). Some points on the back side (opposite to the corner) of yield surfaces are not very well fitted by the new model. In fact, these points are much more difficult to detect precisely: the stress strain curve presents, at the beginning of plasticity, a very progressive curvature in the unloading direction instead of the (easy to detect) sharp transition in the loading direction. Some experimental yield stresses do not comply with the symmetry with respect to the $e_1$ tensorial direction (horizontal axis on Fig. 1) that is expected. This may be due to an initial anisotropy of the material or some other experimental problem.

The second experiment, a non proportional tension-torsion loading, is described in section 2. The theoretical curves are obtained with the values obtained by the identification process described earlier on the previous monotonic tensile testing (Table 1). The average distance (Eqn 4.2) from measured yield surfaces to simulated ones is 11.25 MPa. Considering all the same constants identical but $X_l$ infinite (classical model) the distance is 13.46 MPa. The corresponding yield surfaces have circular representation on Fig. (3), they are not drawn for clarity. The observation made before about back side points for tensile testing remains true, but the error is considerably bigger than in the monotonic case; this leads to the supposition that the isotropic hardening rule, as a monotonic function of r (or p in the classic case, see section 3) overestimates the size of the elasticity domain in the case of non proportional loading. Finally, one can observe that the experimental points seem to be distant from the theoretical curve proportionally to the intensity of the shear stress. This is demonstrated also for the initial (von Mises type) yield surface. The aluminum may obey to a criterion distinguishing the tension's deviator from other shears, *e.g.* Tresca's criterion. But these criteria do not respect Ilyushin's [1954] postulate thus prohibiting the very useful intrinsic writing of plasticity in deviatoric subspace.

Table 1. Material constants identified for the 2024-T4 aluminium alloy.

| $\sigma_y$ (MPa) | C (MPa) | m | k (MPa) | $\gamma$ | $X_l$ (MPa) |
|---|---|---|---|---|---|
| 156 | 11,800 | 1.4 | 331 | 103 | 130 |

## CONCLUSION

The presented theory allows a simple modeling of the distortion of yield surfaces within the classical thermodynamical framework; it has many possible applications such as metal forming, fatigue, forming limit curves, *etc*, and constitutes a generalization of the classical theory. It can be easily adapted to more sophisticated models including, for example, the models improved for ratchetting (Chaboche [1994]), with a reasonable augmentation of calculation time. Some variants can be envisaged: it is possible to link the distortion effect to a variable other than X, for example one of the $X_i$ for multiple kinematic hardening models, or to the plastic strain $e^p$. Some two-dimensional ratchetting experiments are planned as the disortion effect takes on a great importance in this type of loading (mainly through the direction of the normal of the yield surface).